\documentstyle[12pt]{article}
\newcommand{\be}{\begin{equation}}
\newcommand{\ee}{\end{equation}}

\newcommand{\beqa}{\begin{eqnarray}}
\newcommand{\eeqa}{\end{eqnarray}}
\newcommand{\nn}{\nonumber}

\newcommand{\eqref}[1]{(\ref{#1})}

\def\boxit#1{\vbox{\hrule\hbox{\vrule\kern8pt
\vbox{\hbox{\kern8pt}\hbox{\vbox{#1}}\hbox{\kern8pt}}
\kern8pt\vrule}\hrule}}
\def\mathboxit#1{\vbox{\hrule\hbox{\vrule\kern8pt\vbox{\kern8pt
\hbox{$\displaystyle #1$}\kern8pt}\kern8pt\vrule}\hrule}}

\def\IB{\relax\hbox{$\inbar\kern-.3em{\rm B}$}}
\def\IC{\relax\hbox{$\inbar\kern-.3em{\rm C}$}}
\def\ID{\relax\hbox{$\inbar\kern-.3em{\rm D}$}}
\def\IE{\relax\hbox{$\inbar\kern-.3em{\rm E}$}}
\def\IF{\relax\hbox{$\inbar\kern-.3em{\rm F}$}}
\def\IG{\relax\hbox{$\inbar\kern-.3em{\rm G}$}}
\def\IGa{\relax\hbox{${\rm I}\kern-.18em\Gamma$}}
\def\IH{\relax{\rm I\kern-.18em H}}
\def\IK{\relax{\rm I\kern-.18em K}}
\def\IL{\relax{\rm I\kern-.18em L}}
\def\IP{\relax{\rm I\kern-.18em P}}
\def\IR{\relax{\rm I\kern-.18em R}}
\def\IZ{\relax\ifmmode\mathchoice
{\hbox{\cmss Z\kern-.4em Z}}{\hbox{\cmss Z\kern-.4em Z}}
{\lower.9pt\hbox{\cmsss Z\kern-.4em Z}} {\lower1.2pt\hbox{\cmsss
Z\kern-.4em Z}}\else{\cmss Z\kern-.4em Z}\fi}

\def\II{\relax{\rm I\kern-.18em I}}

\def\CL {{\cal L}}

\def\CP {{\cal P}}

\begin{document}

\begin{flushright}
NRCPS-HE-10-01\\
January, 2010
\end{flushright}

\vspace{1cm}
\begin{center}

{\Large \it Topological Mass  Generation\\
\vspace{0,3cm}
in  \\
\vspace{0,3cm}
Four-Dimensional Gauge Theory\\
} 

\vspace{1cm}

{ \it{George  Savvidy  } }

\vspace{0.5cm}

 {\it Institute of Nuclear Physics,} \\
{\it Demokritos National Research Center }\\
{\it Agia Paraskevi, GR-15310 Athens, Greece} 
\end{center}
\vspace{60pt}

\centerline{{\bf Abstract}}

\vspace{12pt}

\noindent
The Lagrangian of non-Abelian tensor gauge fields describes the interaction of
the Yang-Mills and massless tensor  bosons of
increasing helicities. We have found a metric-independent gauge invariant
density  which is a four-dimensional analog of the
Chern-Simons density. The Lagrangian augmented by this Chern-Simons-like invariant
describes massive Yang-Mills boson,
providing a gauge-invariant mass gap for a four-dimensional gauge field theory.
We present invariant densities which can provide masses to
the high rank tensor bosons.

\vspace{150 pt}


\newpage

\pagestyle{plain}

\section{\it Introduction}
Several mechanisms are currently known for generating massive vector particles that
are compatible with the gauge invariance. One of them
is the spontaneous symmetry breaking mechanism, which
generates masses and requires the existence of
the fundamental scalar particle - the Higgs boson. The scalar field
provides the longitudinal polarization of the massive vector boson
and ensures unitarity of its scattering
amplitudes \cite{Cornwall:1973tb,Llewellyn Smith:1973ey}\footnote{
Extended discussion and references can be found in
\cite{Schwinger:1962tn,Schwinger:1962tp,Goto:1967,Veltman:1968ki,Slavnov:1970tk,
vanDam:1970vg,Slavnov:1972tk,Veltman:2000xp,Sikivie:1980fm,
Farhi:1980xs,Dimopoulos:1980fj,Csaki:2003dt,Csaki:2003zu,
Burdman:2003ya,Slavnov:2006rf}.}.

The argument in favor of a pure gauge field theory mechanism
was a dynamical mechanism of mass generation proposed  by Schwinger
\cite{Schwinger:1962tn}, who was arguing that the gauge
invariance of a vector field does not necessarily lead to the massless spectrum
of its excitations and suggested its realization in (1+1)-dimensional gauge
theory\cite{Schwinger:1962tp}.

Compatibility of gauge invariance and mass term in
(2+1)-dimensional gauge field theory was demonstrated by
Deser, Jackiw and Templeton \cite{Deser:1982vy, Deser:1981wh} and
Schonfeld \cite{Schonfeld:1980kb}, who added to
the YM Lagrangian a gauge invariant Chern-Simons density:
\beqa
\CL_{YMCS}= - {1\over 2}Tr G_{ij }G_{ij } + {\mu \over 2}~
\varepsilon_{ijk}  ~Tr~ (A_{i}\partial_{j}A_{k }
-i g {2\over 3} A_{i}A_{j}A_{k}), \nn
\eeqa
where $G_{ij }$ is a field strength tensor.
The mass parameter $\mu$ carries dimension
of $[mass]^1$.
The corresponding free equation of motion
for the vector potential $A_i = e_i e^{ikx}$ has the form
$$
(-k^2 \eta_{ij} +k_i k_j )e_j +i \mu ~\varepsilon_{ijl}~ k_j e_l =0
$$
and  shows that the gauge field excitation becomes massive.

In this article we suggest a similar mechanism that generates masses of the YM boson
and tensor gauge bosons in (3+1)-dimensional space-time at the classical level.
As we shall
see, in  non-Abelian tensor gauge theory \cite{Savvidy:2005fi,Savvidy:2005zm,Savvidy:2005ki}
there exists a gauge invariant,
metric-independent density $\Gamma $ in five-dimensional
space-time\footnote{The definition of the higher-rank field strength tensors is
given by formula (\ref{fieldstrengthparticular}).}:
\be
\Gamma  = \varepsilon_{lmnpq} TrG_{lm}G_{np, q}
= \partial_{l} \Sigma_{l},
\ee
which is the derivative of the vector current $\Sigma_{l}$ (l,..=0,1,...,4).
This invariant in five dimensions has many properties
of the Chern-Pontryagin density $\CP= \partial_{\mu} C_{\mu}$ in
four-dimensional YM theory, which  is
a derivative of the Chern-Simons topological vector current $C_{\mu}$.
Considering the fifth component of the vector
current $\Sigma_{4}  \equiv \Sigma $  and
fields which are independent on the fifth spacial coordinate $x_4$, one can get a
gauge invariant density  which is defined in four-dimensional
space-time\footnote{We are using
Greek letters to numerate four-dimensional coordinates.}:
\be
\Sigma
= \varepsilon_{\mu\nu\rho \lambda}   Tr ~G_{\mu\nu} A_{\rho \lambda } .
\ee
Its dimensionality
is $[mass]^3$, therefore in order to get dimensionless functional in
four dimensions we should multiply
it by the parameter~  $m  $ ~which has dimensionality $[mass]^1$.
Adding this term  to the Lagrangian
of non-Abelian tensor gauge fields leaves intact its gauge
invariance, and to lowest order in coupling constant the equations of  motion for
 the YM field $A_{\mu}=e_{\mu}e^{ikx}$ of helicities $\lambda=\pm 1$
 and for the antisymmetric part $B_{\mu\nu}=b_{\mu\nu}e^{ikx}$ of the
 rank-2 gauge field
$A_{\mu\nu}$, which carries helicity zero $\lambda=0$ state,
can be written in the following form:
\beqa
 (-k^2 \eta_{\nu \mu }+
k_{\nu} k_{\mu} ) e_{\mu }  +
 i m ~\varepsilon_{\nu \mu\lambda\rho} k_{\mu} b_{\lambda\rho}= 0,~~~~~~~~~~~~~~~~\nn\\
(- k^{2} \eta_{\nu \mu } \eta_{\lambda\rho}
+k_{\nu} k_{\mu}\eta_{\lambda\rho}
-\eta_{\nu \mu } k_{\lambda} k_{\mu} ) b_{\mu\rho}  +
i {2 m \over 3}  ~\varepsilon_{\nu\lambda\mu \rho} k_{\mu} e_{\rho} = 0.~~~~\nn
\eeqa
These field equations  describe massive state  of the vector particle of the  mass
\be
M^2 =  {4 \over 3 } ~m^2 .
\ee
Thus at the classical level the YM vector boson becomes massive. The
anti-symmetric tensor $B^a_{\mu\nu}$, which carries zero helicity state, provides
the longitudinal polarization of the massive vector meson,
suggesting an alternative
mechanism for mass generation in non-Abelian gauge field theories in
four-dimensional space-time.
Because both of the fields, the vector and the antisymmetric tensor, are in the adjoint representation, it
follows  that
all vector fields $A^{a}_{\mu}, ~a=1,...,dim G$ acquire the same mass M.
At this stage the symmetric part $A^{S}_{\mu\nu}$ of the rank-2 gauge field, which carries helicities
$\lambda=\pm 2$, remains massless.

As a next step we shall demonstrate that in five-dimensional space-time  there
actually exists an
infinite series of invariants  $\Gamma_s~~(s=1,3,...)$ which are constructed
by means of the
totaly antisymmetric Levi-Civita epsilon tensor $\varepsilon_{lmnpq}$
in combination with the generalized field strength tensors $G_{mn,l_1 ... l_{s}}$.
These invariants can be represented as total derivatives of the vector currents
$\Sigma^{s}_l$:
$$
\Gamma_s = \partial_{l} \Sigma^{s}_{l},
$$
where the vector currents  $\Sigma^{s}_l $ involve a free index $l $ carried
by the Levi-Civita epsilon tensor. Considering the fifth component of the vector
current $\Sigma^{s}_4 \equiv \Sigma_s$ one can see that the
remaining indices will not repeat the external index.
Furthermore, if all dependence
of the tensor gauge fields on the fifth spacial coordinate $x_4$ is suppressed, we are left with the
invariant densities which are defined in four-dimensional space-time.
Their dimensionality
is $[mass]^3$ therefore in order to get dimensionless functional in (3+1)-dimensions we should multiply
them by the parameters $m_s$ which have units $[mass]^1$:
$$
 m_s \int_{M_4} \Sigma_s.
$$
Adding these densities to the Lagrangian of non-Abelian tensor fields keeps intact its gauge
invariance, up to total divergence terms, so that the Lagrangian takes the following form:
\be\label{fulllagrangiantop}
{{\cal L}}_m = {{\cal L}}_{YM} +   \sum_s
({{\cal L}}_{s+1}+ {2s\over s+1}{{\cal L}}^{'}_{s+1})+
\sum_s m_s ~\Sigma_s.
\ee
The natural appearance of the mass parameters
hints at the fact that the theory turns out to be a massive theory.

In the next section we present a short introduction into the theory of non-Abelian tensor
gauge fields defining their gauge transformations, fields strength tensors and the
Lagrangian \cite{Savvidy:2005fi,Savvidy:2005zm,Savvidy:2005ki}.
In section three we derive different properties of the
metric-independent and gauge invariant density $\Gamma $ and of the corresponding vector
current  $\Sigma_l$ and  its reduction to four dimensions.
In section four we overview the helicity content of the massless
tensor gauge fields before including the topological mass term $m \Sigma$
into the Lagrangian. In section five we analyze how
the spectrum of the theory is changing when we add the invariant  $m \Sigma$
to the Lagrangian. As we shall see, a massive spin-1 YM boson with its
three spin polarizations, $\lambda=\pm 1,0$ appears in the spectrum, while
the antisymmetric field has been absorbed as its longitudinal polarization.
In the sixth section we are presenting infinite series of
dimensionful  invariants  $\Sigma_s~~(s=1,3,...)$ which exist in four-dimensional space-time
and can be added to Lagrangian in order to generate masses of the tensor bosons. In
section seven  we are presenting topological invariants in six dimensions.

\section{\it Non-Abelian Gauge Fields}
In our recent approach  the gauge fields are defined as
rank-$(s+1)$ tensors
\cite{Savvidy:2005fi,Savvidy:2005zm,Savvidy:2005ki}
$$
A^{a}_{\mu\lambda_1 ... \lambda_{s}}(x),
$$
which are totally symmetric with respect to the
indices $  \lambda_1 ... \lambda_{s}  $. The number of symmetric
indices $s$ runs from zero to infinity
\footnote{  {\it A priori} the tensor fields
have no symmetries with respect to the first index  $\mu$.
The free field theory of totally symmetric tensors of high rank were constructed in
\cite{fierz,fierzpauli,schwinger,
Weinberg:1964cn,vanDam:1970vg,singh,fronsdal,Bengtsson:1983pd,Guttenberg:2008qe}.}.
The index $a$ numerates the generators $L^a$
of an appropriate Lie algebra.
The extended non-Abelian gauge transformation $\delta_{\xi} $ of the tensor gauge fields
is defined in the Appendix   and comprises a closed algebraic structure.
The generalized field strength tensors
are defined as follows
\cite{Savvidy:2005fi,Savvidy:2005zm,Savvidy:2005ki}:
\beqa\label{fieldstrengthparticular}
G_{\mu\nu} &=&
\partial_{\mu} A_{\nu} - \partial_{\nu} A_{\mu} -
i g [A_{\mu}~A_{\nu}],\\
G_{\mu\nu,\lambda} &=&
\partial_{\mu} A_{\nu\lambda} - \partial_{\nu} A_{\mu\lambda} -
i g  (~[A_{\mu}~A_{\nu\lambda}] + [A_{\mu\lambda}~A_{\nu}] ~),\nn\\
G_{\mu\nu,\lambda\rho} &=&
\partial_{\mu} A_{\nu\lambda\rho} - \partial_{\nu} A_{\mu\lambda\rho} -
i g (~[A_{\mu}~A_{\nu\lambda\rho}] +
 [A_{\mu\lambda}~A_{\nu\rho}]+[A_{\mu\rho}~A_{\nu\lambda}]
 + [A_{\mu\lambda\rho}~A_{\nu}] ~),\nn\\
 ......&.&............................................\nn
\eeqa
and transform homogeneously
with respect to the extended gauge transformations $\delta_{\xi} $.
The tensor gauge fields are in the matrix representation
$A^{ab}_{\mu\lambda_1 ... \lambda_{s}} =
(L_c)^{ab}  A^{c}_{\mu\lambda_1 ... \lambda_{s}} = i f^{acb}A^{c}_{\mu
\lambda_1 ... \lambda_{s}}$  and
$f^{abc}$ are the structure constants of the Lie algebra.

Using field strength tensors one can construct two infinite series of forms
$
{{\cal L}}_{s}$ and ${{\cal L}}^{'}_{s}~
$
invariant with respect to the
transformations $\delta_{\xi} $. They are quadratic in field strength
tensors. The
first series is given by the formula \cite{Savvidy:2005fi,Savvidy:2005zm,Savvidy:2005ki}
\beqa\label{fulllagrangian1}
{{\cal L}}_{s+1}
&=& -{1\over 4}(\sum^{2s}_{i=0}~a^{s}_i ~
G^{a}_{\mu\nu, \lambda_1 ... \lambda_i}~
G^{a}_{\mu\nu, \lambda_{i+1}...\lambda_{2s}})
(\sum_{P } \eta^{\lambda_{i_1} \lambda_{i_2}} .......
\eta^{\lambda_{i_{2s-1}} \lambda_{i_{2s}}})~,
\eeqa
where the sum $\sum_P$ runs over all nonequal permutations of
$\lambda_i~'s$ and  $a^{s}_i = {s!\over i!(2s-i)!}$.
The second series of gauge invariant quadratic forms is given by the formula
\cite{Savvidy:2005fi,Savvidy:2005zm,Savvidy:2005ki,Savvidy:2005vm,Savvidy:2009zz}
\beqa\label{secondfulllagrangian}
{{\cal L}}^{'}_{s+1}
&=& {1\over 4}(\sum^{2s+1}_{i=1}~{ a^{s}_{i-1}\over s}  ~
G^{a}_{\mu\lambda_1,\lambda_2  ... \lambda_i}~
G^{a}_{\mu\lambda_{i_{2s+2}},\lambda_{i+1}...\lambda_{2s+1}})
(\sum^{'}_{P} \eta^{\lambda_{i_1} \lambda_{i_2}} .......
\eta^{\lambda_{i_{2s+1}} \lambda_{i_{2s+2}}})~,
\eeqa
where the sum $\sum^{'}_P$ runs over all nonequal permutations of
$\lambda_i~'s$, with exclusion
of the terms which contain $\eta^{\lambda_{1},\lambda_{2s+2}}$.

These forms
contain quadratic   kinetic terms, as well as cubic and
quartic terms  describing
nonlinear interaction of gauge fields with dimensionless
coupling constant $g$.
In order to make all tensor gauge fields dynamical one should add all these
forms in the Lagrangian
\cite{Savvidy:2005fi,Savvidy:2005zm,Savvidy:2005ki,Savvidy:2005vm,Savvidy:2009zz}:
\be\label{fulllagrangian3}
{{\cal L}} = {{\cal L}}_{YM} +  {{\cal L}}_{2}+ {{\cal L}}^{'}_{2}
+ ...+
g_{s+1}({{\cal L}}_{s+1}+ {2s\over s+1}{{\cal L}}^{'}_{s+1})+....
\ee
The coupling constants  $g_{3},  g_4,... $ remain arbitrary
because each term is separately invariant
with respect to the extended gauge transformations $\delta_\xi$
and still leaves these coupling constants undetermined. The Lagrangian
$\CL$ is well defined in any dimension.

\section{\it Metric-Independent Density $\Gamma$}

Let us consider a new invariant in five-dimensional space-time $(4+1)$,
which can be constructed by means of the
totaly antisymmetric Levi-Civita epsilon tensor $\varepsilon_{\mu\nu\lambda\rho\sigma}$
($\mu,\nu,...=0,1,2,3,4$) in combination with the generalized field strength tensors
\beqa\label{freeactionthreeprimesum}
\Gamma
= \varepsilon_{\mu\nu\lambda\rho\sigma} TrG_{\mu\nu}G_{\lambda\rho, \sigma}
=2~\varepsilon_{\mu\nu\lambda\rho\sigma}G^{a}_{\mu\nu}G^{a}_{\lambda\rho, \sigma}.
\eeqa
We shall demonstrate that this invariant in five dimensions has many properties
of the Chern-Pontryagin density
\be\label{chernpontragyn}
\CP= {1\over 4}\varepsilon_{\mu\nu\lambda\rho }Tr G_{\mu\nu}G_{\lambda\rho }=\partial_{\mu} C_{\mu}
\ee
in Yang-Mill theory in four dimensions, where
\be\label{chernsimons}
C_{\mu}=\varepsilon_{\mu\nu\lambda\rho }Tr (A_{\nu}\partial_{\lambda}A_{\rho }
-i{2\over 3}g  A_{\nu}A_{\lambda}A_{\rho })
\ee
is the Chern-Simons topological current.
Indeed, $\Gamma$ is obviously diffeomorphism-invariant and does not involve a
space-time metric.
It is gauge invariant because under the gauge transformation $\delta_{\xi} $
(\ref{fieldstrengthtensortransfor}) it vanishes:
\beqa
\delta_{\xi} \Gamma &=& \varepsilon_{\mu\nu\lambda\rho\sigma} Tr (\delta G_{\mu\nu} G_{\lambda\rho, \sigma}+
G_{\mu\nu} \delta G_{\lambda\rho, \sigma}) \nn\\
&=&-i g  \varepsilon_{\mu\nu\lambda\rho\sigma} Tr ( [G_{\mu\nu} ~\xi] G_{\lambda\rho, \sigma}
 +G_{\mu\nu} (~[~G_{\lambda\rho,\sigma}~ \xi ]
+  [G_{\lambda\rho} ~\xi_{\sigma}]~))=0.\nn
\eeqa
The variation of its integral over the gauge fields
$A^{a}_{\mu}$ and $A^{a}_{\mu\lambda}$ gives:
\beqa\label{topologicalinvariant}
\delta_{A} \int_{M_5} d^5x ~\Gamma &=&\varepsilon_{\mu\nu\lambda\rho\sigma} \int d^5x Tr
((\nabla_{\mu} \delta A_{\nu}-
\nabla_{\nu} \delta A_{\mu}) G_{\lambda\rho, \sigma}+\nn\\
&+&
G_{\mu\nu}(\nabla_{\lambda} \delta A_{\rho\sigma}-
\nabla_{\rho} \delta A_{\lambda\sigma}  -i g [\delta A_{\lambda} ~A_{\rho\sigma}]
-i g [\delta A_{\lambda \rho} ~\delta A_{\rho}]))=\nn
\\
&-&  2\varepsilon_{\mu\nu\lambda\rho\sigma} \int d^5x Tr
((\nabla_{\mu} G_{\lambda\rho, \sigma}
- i g [A_{\mu\sigma} ~G_{\lambda\rho}])~\delta A_{\nu}
+(\nabla_{\lambda} G_{\mu\nu}) ~\delta A_{\rho\sigma})\nn\\
&+& 2 \varepsilon_{\mu\nu\lambda\rho\sigma} \int d^5x Tr
(\nabla_{\mu} (G_{\lambda\rho, \sigma} ~\delta A_{\nu})
+\nabla_{\lambda} (G_{\mu\nu} ~\delta A_{\rho\sigma})) .\nn
\eeqa
Recalling the Bianchi identity in YM theory
and the generalized Bianchi identities for higher-rank  field strength tensor $G_{\nu\lambda,\rho}$
presented in the Appendix, one can see that  $\Gamma$  gets contribution
only from the boundary terms and vanishes when the fields vary in the
bulk of the manifold\footnote{The trace of the commutators vanishes:
$Tr([A_{\mu} ;G_{\lambda\rho, \sigma} \delta A_{\nu}]
+[A_{\lambda};  G_{\mu\nu} \delta A_{\rho\sigma}])=0$.}:
\beqa
\delta_{A} \int_{M_5} d^5x ~\Gamma =  2 \varepsilon_{\mu\nu\lambda\rho\sigma}
\int_{M_5}  d^5x ~\partial_{\mu} ~Tr
(G_{\lambda\rho, \sigma}~\delta A_{\nu}
+G_{\nu\lambda} ~\delta A_{\rho\sigma})=\nn
\\
= 2 \varepsilon_{\mu\nu\lambda\rho\sigma}
\int_{\partial M_5}~Tr
(G_{\lambda\rho, \sigma}~\delta A_{\nu}
+G_{\nu\lambda} ~\delta A_{\rho\sigma}) d\sigma_{\mu} =0.\nn
\eeqa
Therefore  $\Gamma$  is insensitive
to the {\it local} variation of the fields.
It became obvious that $\Gamma$ is a total derivative of some
vector current $\Sigma_{\mu}$. Indeed, simple algebraic computation gives
\beqa\label{exactform}
\Gamma
=\varepsilon_{\mu\nu\lambda\rho\sigma} TrG_{\mu\nu}G_{\lambda\rho, \sigma}=
\partial_{\mu} \Sigma_{\mu},
\eeqa
where
\beqa\label{topologicalcurrent}
\Sigma_{\mu}
&=& 2 \varepsilon_{\mu\nu\lambda\rho\sigma} Tr (A_{\nu} ~\partial_{\lambda} A_{\varrho\sigma }
-  \partial_{\lambda} A_{\nu} ~ A_{\rho\sigma }
- 2 i g A_{\nu} A_{\lambda} A_{\rho\sigma }).
\eeqa
After some rearrangement and taking into account the definition of the field strength
tensors (\ref{fieldstrengthparticular}) we can get the following
form of the vector current:
\beqa\label{topologicalcurrent1}
\Sigma_{\mu}
&=&  \varepsilon_{\mu\nu\lambda\rho\sigma} Tr G_{\nu\lambda }A_{\rho  \sigma}.
\eeqa
It is instructive  to compare the expressions (\ref{freeactionthreeprimesum}), (\ref{chernpontragyn})
and (\ref{chernsimons}), (\ref{topologicalcurrent1}). Both entities $\CP$
and $\Gamma$ are
metric-independent, are insensitive to the local variation of the fields and are derivatives of the
corresponding vector currents $C_{\mu}$ and $\Sigma_{\mu}$. The difference between them is that the former is
defined in four dimensions, while the latter in five. This difference in one unit of the space-time
dimension originates from the fact that we have at our disposal high-rank tensor gauge fields
to build new invariants.
The same is true for the Chern-Simons topological current $C_{\mu}$ and for
the current $\Sigma_{\mu}$, where
the latter is defined in five dimensions.
It is also remarkable that the current  $\Sigma_{\mu}$ is linear in YM field strength tensor
and in the rank-2 gauge field, picking up only its antisymmetric part.

While the invariant $\Gamma$ and the vector current $\Sigma_{\mu}$ are defined on a five-dimensional
manifold, we may restrict the latter to one lower, four-dimensional manifold. The
restriction proceeds as follows.
Let us consider the fifth component of the vector current $\Sigma_{\mu}$:
\beqa\label{topologicalcharge}
\Sigma \equiv \Sigma_{4}
&=&  \varepsilon_{4\nu\lambda\rho\sigma} Tr G_{\nu\lambda }A_{\rho  \sigma}.
\eeqa
Considering the fifth component of the vector
current $\Sigma \equiv\Sigma_4$ one can see that the remaining indices will not repeat
the external index and the sum is restricted to the sum over indices of four-dimensional
space-time.
Therefore we can reduce this functional to four dimensions.
This is the case when the gauge fields are independent
on the fifth coordinate $x_4$. Thus the density $\Sigma$ is well
defined in four-dimensional space-time and, as we shall see,
it is also gauge invariant up to the total divergence term.  Therefore we shall
consider its integral over four-dimensional space-time\footnote{Below we
are using the same
Greek letters to numerate now the four-dimensional coordinates. There should be
no confusion because the
dimension  can always be recovered from  the dimension of the epsilon tensor.}:
\beqa\label{topologicalSigmaCS}
\int_{M_4} d^4 x~ \Sigma
&=&  \varepsilon_{\nu\lambda\rho\sigma} \int_{M_4} d^4 x~ Tr~ G_{\nu\lambda } A_{\rho  \sigma} .
\eeqa
This entity is an analog of the Chern-Simon secondary characteristic
\be\label{chernsimonscharcterictic}
CS =\varepsilon_{ijk } \int_{M_3}  d^3x ~Tr~ (A_{i}\partial_{j}A_{k }
-i g {2\over 3} A_{i}A_{j}A_{k }),
\ee
but, importantly, instead of being defined in three dimensions
it is now  defined in four dimensions. Thus the non-Abelian tensor gauge fields
allow to build a natural generalization of the Chern-Simons characteristic in four-dimensional
space-time.

As we claimed this functional is gauge invariant up to the total divergence term.
Indeed, its gauge variation under $\delta_\xi$ (\ref{polygauge}),
(\ref{fieldstrengthtensortransfor}) is
\beqa\label{variation}
\delta_{\xi} \int_{M_4} d^4 x~ \Sigma
&=& \varepsilon_{\nu\lambda\rho\sigma} \int_{M_4} Tr (-i g [G_{\nu\lambda }~\xi] A_{\rho  \sigma}
+ G_{\nu\lambda } (\nabla_{\rho} \xi_{\sigma}-i g  [A_{\rho  \sigma}~\xi] )) d^4 x =\nn\\
&=&\varepsilon_{\nu\lambda\rho\sigma} \int_{M_4} \partial_{\rho} ~Tr (
G_{\nu\lambda }  \xi_{\sigma} ) d^4 x=\varepsilon_{\nu\lambda\rho\sigma}
\int_{\partial M_4} Tr (
G_{\nu\lambda }  \xi_{\sigma} ) d\sigma_{\rho} =0 .
\eeqa
Here the first and the third  terms cancel each other and the second one, after integration by
part and recalling the Bianchi identity  (\ref{newbianchi}),  leaves only the
boundary term, which vanishes  when
the gauge parameter $\xi_{\sigma}$ tends to zero at infinity.

It is interesting to know whether or not the invariant $\Sigma$ is associated with some
new topological characteristic of the gauge fields.
If the YM field strength $G_{\nu\lambda }$ vanishes, then the vector potential is equal to the
pure gauge connection $A_{\mu}=U^{-}\partial_{\mu} U$. Inspecting the expression for the
invariant $\Sigma$ one can get convinced that it vanishes on such fields because there is
a field strength tensor $G_{\nu\lambda }$ in the integrant. Therefore
it does not differentiate topological properties of the gauge function $U$,
like its winding number. Both "small" and "large" gauge transformations have zero
contribution to this invariant. It may distinguish fields which are
falling less faster at infinity and have nonzero field strength tensor $G_{\nu\lambda }$
and the tensor gauge field $A_{\rho  \sigma} $.

The dimension of this functional
is not difficult to calculate. In four dimensions the gauge fields have dimension of $[mass]^1$,
therefore if we intend to add this new density to the Lagrangian we should introduce the mass parameter $m$:
\beqa\label{topologicalmass}
m~\Sigma
&=&  m~\varepsilon_{\nu\lambda\rho\sigma}   Tr ~G_{\nu\lambda } A_{\rho  \sigma} ,
\eeqa
where parameter $m$ has units $[mass]^1$. Adding this term  to the Lagrangian
of non-Abelian tensor gauge fields keeps intact its gauge
invariance and our aim is to analyze the particle spectrum of this gauge field theory.
The natural appearance of the mass parameters
hints at the fact that the theory turns out to be a massive theory. We shall see
that the YM vector boson becomes massive, suggesting an alternative
mechanism for mass generation
in gauge field theories in four-dimensional space-time.

We have to notice that the Abelian version of the invariant $\Sigma$ was investigated earlier
in \cite{Cremmer:1973mg,Kalb:1974yc,Nambu:1975ba,Aurilia:1981xg,
Freedman:1980us,Slavnov:1988sj,Allen:1990gb,Govindarajan:1981jp,
Lahiri:1996dm,Deguchi:2008by,Dvali:2005ws}.
Indeed, if one considers instead of a non-Abelian group the Abelian group one can see
that the invariant $\Sigma$ reduces to the
$\varepsilon_{\nu\lambda\rho\sigma}   F_{\nu\lambda } B_{\rho  \sigma}$ and when
added to the Maxwell Lagrangian provides a  mass to the vector field
\cite{Cremmer:1973mg,Kalb:1974yc,Ogievetsky:1967ij,Nambu:1975ba,Aurilia:1981xg,
Freedman:1980us,Aldaya:2009qb}.
Attempts at producing a non-Abelian  invariant
in a similar way  have come up with difficulties  because they
involve non-Abelian generalization of
gauge transformations of antisymmetric fields
\cite{Freedman:1980us,Allen:1990gb,Leblanc:1991av}. Let us compare the
formulas (2.16) and (2.17) suggested in \cite{Freedman:1980us,Slavnov:1988sj}
for the transformation of antisymmetric field with the gauge transformation
$\delta_{\xi}$  (\ref{polygauge}).
For lower-rank fields the latter can be written in the following way
\cite{Savvidy:2005fi,Savvidy:2005zm,Savvidy:2005ki}:
\beqa
\delta_{\xi}  A_{\mu} &=& \partial_{\mu}\xi -i g[A_{\mu},\xi],~~~~~~~~~~~~~~~~~~~
\delta_{\zeta}  A_{\mu} = 0,
\nonumber\\
\delta_{\xi}  A_{\mu\nu} &=& -i g [A_{\mu\nu},\xi],~~~~~~~~~~~~~~~~~~~~~~~
\delta_{\zeta}  A_{\mu\nu} = \partial_{\mu}\zeta_{\nu} -i g[A_{\mu},\zeta_{\nu}] .\nonumber
\eeqa
The  {\it antisymmetric part} of this transformation coincides with the one
suggested  in \cite{Freedman:1980us} if one takes also the auxiliary field $A^i_{\mu}$ of
\cite{Freedman:1980us} equal to zero.
The crucial point is that the gauge transformations of non-Abelian tensor gauge fields
$\delta_{\xi}$   (\ref{polygauge})  cannot be limited to a YM vector and antisymmetric
field $B^{a}_{\mu\nu}$. Instead, antisymmetric field is
augmented by a symmetric rank-2 gauge field, so that together they form a
gauge field $A^{a}_{\mu\nu}$ which transforms as it is given above and is a fully propagating
field.
It is also important that one should include all high-rank gauge fields in order to be able
to close the group of gauge transformations and to construct
invariant Lagrangian (\ref{fulllagrangian3}).

Let us shortly overview the helicity content of the massless tensor gauge fields
before including a massive term  into the
Lagrangian \cite{Savvidy:2005fi,Savvidy:2005zm,Savvidy:2005ki,
Konitopoulos:2007hw,Savvidy:2009zz}.

\section{\it Helicity Content of Massless Tensor Gauge Fields}

Let us first recapitulate the analysis of the
particle spectrum before including new massive terms into the Lagrangian
\cite{Savvidy:2005fi,Savvidy:2005zm,Savvidy:2005ki}.
In the Yang-Mills theory
$$
\CL_{YM} = - {1\over
4}G^{a}_{\mu\nu }G^{a}_{\mu\nu }\nn
$$
the free equation of motion is
$$
\partial_{\mu} F^{a}_{\mu\nu }=0,
$$
or, in terms of vector gauge field,
$$
(\eta_{\mu\nu} \partial^{2} -
 \partial_{\mu} \partial_{\nu} )
A^{a}_{\nu}=0,
$$
and it describes the propagation of massless gauge boson of helicity $\lambda = \pm 1$.

The second term in (\ref{fulllagrangian3}) of the Lagrangian $\CL$ defines  the
kinetic operator and the interactions of the rank-2 gauge field
$A^{a}_{\mu\lambda}$\footnote{It has sixteen components in the
four-dimensional space-time.}:
\beqa\label{totalactiontwo}
{{\cal L}}_2 +  {{\cal L}}^{'}_2  =
&-&{1\over 4}G^{a}_{\mu\nu,\lambda}G^{a}_{\mu\nu,\lambda}
-{1\over 4}G^{a}_{\mu\nu}G^{a}_{\mu\nu,\lambda\lambda}\\
&+&{1\over 4}G^{a}_{\mu\nu,\lambda}G^{a}_{\mu\lambda,\nu}
+{1\over 4}G^{a}_{\mu\nu,\nu}G^{a}_{\mu\lambda,\lambda}
+{1\over 2}G^{a}_{\mu\nu}G^{a}_{\mu\lambda,\nu\lambda}.\nn
\eeqa
Its free equation of motion is \cite{Savvidy:2005fi,Savvidy:2005zm,Savvidy:2005ki}
\be
 \partial_{\mu} F^{a}_{\mu\nu,\lambda}
-{1\over 2} (\partial_{\mu} F^{a}_{\mu\lambda,\nu}
+\partial_{\mu} F^{a}_{\lambda\nu,\mu}
+\partial_{\lambda}F^{a}_{\mu\nu,\mu}
+\eta_{\nu\lambda} \partial_{\mu}F^{a}_{\mu\rho,\rho}) = 0,
\ee
where $F^{a}_{\mu\nu,\lambda} = \partial_{\mu} A^{a}_{\nu \lambda} -
\partial_{\nu} A^{a}_{\mu \lambda}$, or, in terms of tensor gauge field,
 it takes the form
\beqa\label{mainequation}
\partial^{2}(A^{a}_{\nu\lambda} -{1\over 2}A^{a}_{\lambda\nu})
-\partial_{\nu} \partial_{\mu}  (A^{a}_{\mu\lambda}-
{1\over 2}A^{a}_{\lambda\mu} ) -
\partial_{\lambda} \partial_{\mu}  (A^{a}_{\nu\mu} - {1\over 2}A^{a}_{\mu\nu} )
+\nn\\
+\partial_{\nu} \partial_{\lambda} ( A^{a}_{\mu\mu}-{1\over 2}A^{a}_{\mu\mu})
 + {1\over 2}\eta_{\nu\lambda} ( \partial_{\mu} \partial_{\rho}A^{a}_{\mu\rho}
-  \partial^{2}A^{a}_{\mu\mu})=0~~~~~~~~~
\eeqa
and is invariant with respect to the group of gauge transformations
\be\label{largegaugetransformation}
\delta A^{a}_{\mu \lambda} =\partial_{\mu} \xi^{a}_{\lambda}+
\partial_{\lambda} \zeta^{a}_{\mu},
\ee
where $\xi^{a}_{\lambda}$ and $\zeta^{a}_{\mu}$ are gauge parameters.
This free equation describes the propagation of massless modes of
{\it helicity-two and helicity-zero, $\lambda = \pm 2, 0$, charged gauge bosons}
\cite{Savvidy:2005fi,Savvidy:2005zm,Savvidy:2005ki,
Konitopoulos:2007hw,Savvidy:2009zz}.
This can be seen by decomposition of the rank-2 gauge
field into symmetric $A^S_{\mu\lambda}$ and antisymmetric parts $B_{\mu\lambda}$.
For the symmetric tensor gauge fields $A^{S}_{\nu\lambda}$ the equation
reduces to the free Einstein and Fierz-Pauli equation
\be\label{freeequationsymmetric}
\partial^{2} A^S_{\nu\lambda}
-\partial_{\nu} \partial_{\mu}  A^S_{\mu\lambda} -
\partial_{\lambda} \partial_{\mu}  A^S_{\mu\nu}
+ \partial_{\nu} \partial_{\lambda}  A^S_{\mu\mu}
+\eta_{\nu\lambda}  (\partial_{\mu} \partial_{\rho}A^S_{\mu\rho}
- \partial^{2} A^S_{\mu\mu}) =0,
\ee
which describes the propagation of massless gauge
boson of helicity two, $\lambda = \pm 2$.
For the antisymmetric part of the tensor field it
reduces to the equation \cite{Ogievetsky:1967ij,Cremmer:1973mg,Kalb:1974yc}
$$
\partial^{2} B_{\nu\lambda}
-\partial_{\nu} \partial_{\mu}  B_{\mu\lambda} +
\partial_{\lambda} \partial_{\mu}  B_{\mu\nu} =0
$$
and describes the propagation of helicity-zero state, $\lambda = 0$. Let us now see
how the spectrum is changing when we add new invariant  $\Sigma$
to the Lagrangian.

\section{\it Particle Spectrum with Topological Mass Term}
With the new mass term we have to consider the Lagrangian
\beqa
\CL  = \CL_{YM} +    \CL_{2}+ \CL^{'}_{2}  + {m \over 4} ~\Sigma~
= ~- {1\over 4}G^{a}_{\mu\nu }G^{a}_{\mu\nu }-~~~~~~~~~~~~~~~~~~~~~~\nn\\
-{1\over 4}G^{a}_{\mu\nu,\lambda}G^{a}_{\mu\nu,\lambda}
 + {1\over 4}G^{a}_{\mu\nu,\lambda}G^{a}_{\mu\lambda,\nu}
+{1\over 4}G^{a}_{\mu\nu,\nu}G^{a}_{\mu\lambda,\lambda}
-{1\over 4}G^{a}_{\mu\nu}G^{a}_{\mu\nu,\lambda\lambda}
+{1\over 2}G^{a}_{\mu\nu}G^{a}_{\mu\lambda,\nu\lambda}+\nn\\
 + {m \over 2}  ~\varepsilon_{\nu\lambda\rho\sigma}
G^{a}_{\nu\lambda }A^{a}_{\rho  \sigma}~~.~~~~~~~~~~~~~~~~~~~~~~~~~~~~~~~~~~~~
\eeqa
The equations of motion  which follow for the YM and rank-2 gauge fields
are\footnote{At this stage we
keep only YM and rank-2 gauge fields in the field equations, the rank-3 gauge
field is inessential for our analysis of the mass spectrum of the lower-rank
gauge fields. In the
next section we shall include higher-rank gauge fields as well. }:
\beqa
 \nabla^{ab}_{\mu}G^{b}_{\mu\nu} + {m \over 2} ~\varepsilon_{\nu \mu\lambda\rho} G^{a}_{\mu\lambda,\rho}
 + ~~~~~~~~~~~~~~~~~~~~~~~~~~~~~~~~ \nn\\
+ g f^{abc} A^{b}_{\mu\lambda} G^{c}_{\mu\nu,\lambda}
-{1\over 2 }g f^{abc} (A^{b}_{\mu\lambda} G^{c}_{\mu\lambda,\nu}
+A^{b}_{\lambda\mu} G^{c}_{\mu\nu,\lambda}
+A^{b}_{\mu\nu} G^{c}_{\mu\lambda,\lambda}
-A^{b}_{\lambda\lambda} G^{c}_{\mu\nu,\mu})
=0,\nn\\
 \nabla^{ab}_{\mu}G^{b}_{\mu\nu,\lambda} +{m \over 2} ~\varepsilon_{\nu\lambda\mu \rho} G^{a}_{\mu\rho}
 -~~~~~~~~~~~~~~~~~~~~~~~~~~~~~~\nn\\
-{1\over 2} (\nabla^{ab}_{\mu}G^{b}_{\mu\lambda,\nu}
+\nabla^{ab}_{\mu}G^{b}_{\lambda\nu,\mu}
+\nabla^{ab}_{\lambda}G^{b}_{\mu\nu,\mu}
+\eta_{\nu\lambda} \nabla^{ab}_{\mu}G^{b}_{\mu\rho,\rho})+~~~~~~~~~~~~~\nn\\
+g f^{abc} A^{b}_{\mu\lambda} G^{c}_{\mu\nu} +
{1\over 2}g f^{abc}(A^{b}_{\mu\nu} G^{c}_{\mu\lambda}
+A^{b}_{\lambda\mu} G^{c}_{\mu\nu}
+A^{b}_{\mu\mu} G^{c}_{\lambda\nu}
-\eta_{\nu\lambda}  A^{b}_{\mu\rho} G^{c}_{\mu\rho})  =0.
\eeqa
Let us consider the free equations when the coupling constant g is equal to zero:
\beqa\label{freeequationmassive}
 \partial_{\mu} F^{a}_{\mu\nu} +
{m \over 2} ~\varepsilon_{\nu \mu\lambda\rho} F^{a}_{\mu\lambda,\rho}= 0,~~~~~~~~~~~~~~~~~~~~~~~~~~~~~~~~\\
 \partial_{\mu} F^{a}_{\mu\nu,\lambda}
-{1\over 2} (\partial_{\mu} F^{a}_{\mu\lambda,\nu}
+\partial_{\mu} F^{a}_{\lambda\nu,\mu}
+\partial_{\lambda}F^{a}_{\mu\nu,\mu}
+\eta_{\nu\lambda} \partial_{\mu}F^{a}_{\mu\rho,\rho})+
 {m \over 2} ~\varepsilon_{\nu\lambda\mu \rho}  F^{a}_{\mu\rho}= 0,\nn
\eeqa
where $F^{a}_{\mu\nu} =  \partial_{\mu} A^{a}_{\nu  } -
\partial_{\nu} A^{a}_{\mu },~
F^{a}_{\mu\nu,\lambda} = \partial_{\mu} A^{a}_{\nu \lambda} -
\partial_{\nu} A^{a}_{\mu \lambda}$.
This is a coupled system of equations which involved the vector YM field
and antisymmetric part of the rank-2 gauge field. This form of the equations
clearly demonstrates why in non-Abelian tensor gauge field theory it is possible
to have equations in four dimensions which include gauge invariant mass term.
In the first equation the free index $\nu$ is attached to the epsilon tensor
and its last three indices are contracted to the field strength tensor of
rank-2 gauge field $F^{a}_{\mu\lambda,\rho}$. Obviously in vector gauge theory
there are no such objects to contract indices in four dimensions.
In the second equation the
free indices $\nu,\lambda$ are attached to the epsilon tensor
and its last two indices are contracted to the YM field strength tensor
$F^{a}_{\mu\rho}$.

In order to analyze the particle content of the free equations
(\ref{freeequationmassive}) it is convenient to derive them in terms
of dual field strength tensors
\be\label{dualstregths}
F^{*}_{\mu\nu}= {1\over 2} \varepsilon_{\mu\nu\lambda\rho}F_{\lambda\rho},~~~~
F^{*}_{\mu }= {1\over 2} \varepsilon_{\mu\nu\lambda\rho}F_{\nu\lambda,\rho},
\ee
so that the equations will take the form\footnote{Thus
the second free equation
can be written in terms of the dual field strength tensor $F^{* }_{\rho}$.
Notice that the full interacting Lagrangian $\CL_2 +\CL^{'}_2$ cannot be written
in terms of this dual tensor
because of the term ${1\over 4}G^{a}_{\mu\nu,\nu}G^{a}_{\mu\lambda,\lambda}$.}
\beqa\label{dualfreeequationmassive}
 {1\over 2} \varepsilon_{\nu \mu \lambda\rho} \partial_{\mu} F^{* a}_{\lambda\rho} -
 m~ F^{*a }_{\nu}&=& 0,\nn\\
{3\over 2} \varepsilon_{\nu\lambda\mu \rho} \partial_{\mu} F^{*a}_{\rho}
- 2 m ~ F^{*a}_{\nu\lambda}&=& 0.
\eeqa
From the first equation it follows that
$$
 \partial_{\mu} F^{* a}_{\lambda\rho} +
 \partial_{\lambda} F^{* a}_{\rho\mu} + \partial_{\rho} F^{* a}_{\mu\lambda}  -
 m~ \varepsilon_{\mu \lambda\rho\nu } F^{* a}_{\nu}= 0,
$$
then taking derivative and using Bianchi identity $\partial_{\mu} F^{*a}_{\mu\nu}=0$  yields
$$
 \partial^2 F^{* a}_{\lambda\rho}  -
 m~ \varepsilon_{\lambda\rho\mu \nu } \partial_{\mu} F^{* a}_{\nu}= 0,
$$
and using the second equation in (\ref{dualfreeequationmassive}) we can get
\be
(\partial^2 + {4 \over 3} m^2)  F^{* a}_{\lambda\rho}  = 0 ,\nn
\ee
which describes a particle  of the  mass
\be\label{massshell}
k^2 =  {4 \over 3 } ~m^2 = M^2.
\ee
In a similar manner from the second equation it follows that
$$
 \partial_{\mu} F^{*a}_{\nu}- \partial_{\nu} F^{*a}_{\mu}
+ {2\over 3}m ~ \varepsilon_{\mu \nu\lambda\rho} F^{*a}_{\lambda\rho}= 0,
$$
then taking derivative and using Bianchi identity $\partial_{\mu} F^{*a}_{\mu}=0$  yields
$$
\partial^2 F^{*a}_{\nu}
- {2\over 3} m ~ \varepsilon_{\nu \mu \lambda\rho} \partial_{\mu}F^{*a}_{\lambda\rho}= 0,
$$
and using the first equation in (\ref{dualfreeequationmassive}) we can get the same result:
\be
 (\partial^2 +
{4 \over 3} m^2) ~ F^{*a}_{\nu}  = 0.
\ee
The above consideration does not tell us how many propagating modes describes the
system of equations (\ref{system}) on the mass-shell (\ref{massshell}).
In order to understand better the structure and the number of propagating modes
one should analyze the corresponding free
equations (\ref{freeequationmassive}) in terms of  fields
\beqa
 \partial^2 A^{a}_{\nu  } -
\partial_{\nu} \partial_{\mu} A^{a}_{\mu }  +
m ~\varepsilon_{\nu \mu\lambda\rho} \partial_{\mu} A^{a}_{\lambda\rho}= 0,~~~~~~~~~~~~~~~~~~~~~\\
 \partial^{2}(A^{a}_{\nu\lambda} -{1\over 2}A^{a}_{\lambda\nu})
-\partial_{\nu} \partial_{\mu}  (A^{a}_{\mu\lambda}-
{1\over 2}A^{a}_{\lambda\mu} )-
\partial_{\lambda} \partial_{\mu}  (A^{a}_{\nu\mu} - {1\over 2}A^{a}_{\mu\nu} )
+~~~~\nn\\
 +\partial_{\nu} \partial_{\lambda} ( A^{a}_{\mu\mu}-{1\over 2}A^{a}_{\mu\mu})
+{1\over 2}\eta_{\nu\lambda} ( \partial_{\mu} \partial_{\rho}A^{a}_{\mu\rho}
-  \partial^{2}A^{a}_{\mu\mu}) +
 m ~\varepsilon_{\nu\lambda\mu \rho} \partial_{\mu} A^{a}_{\rho} = 0.\nn
\eeqa
As we already mentioned, only the antisymmetric part $B_{\nu\lambda}$
of the rank-2 gauge field $A_{\nu\lambda}$ interacts through the
mass term, the symmetric part $A^S_{\nu\lambda}$ completely decouples from both equations, therefore
we arrive at the following system of equations:
\beqa\label{system}
 \partial^2 A_{\nu  } -
\partial_{\nu} \partial_{\mu} A_{\mu }  +
 m ~\varepsilon_{\nu \mu\lambda\rho} \partial_{\mu} B_{\lambda\rho}= 0,~~~~~~~~~~\nn\\
 \partial^{2} B_{\nu\lambda}
-\partial_{\nu} \partial_{\mu}  B_{\mu\lambda} +
\partial_{\lambda} \partial_{\mu} B_{\mu\nu}  +
{2 m \over 3}  ~\varepsilon_{\nu\lambda\mu \rho} \partial_{\mu} A_{\rho} = 0.
\eeqa
The symmetric part $A^S_{\nu\lambda}$  fulfils the massless equation (\ref{freeequationsymmetric}) and
therefore it is not influenced by the given mass
term\footnote{As we shall see in the next section the symmetric field
can acquire a mass when we include the next invariant mass term $m_3~\Sigma_3$.}.

One can find the structure and the
number of propagating modes calculating the rank of the system (\ref{system})
when it is written in the momentum
representation\footnote{We are using the method developed in
\cite{Konitopoulos:2007hw,Savvidy:2009zz}.}:
\beqa\label{systeminmomentum}
 (-k^2 \eta_{\nu \mu }+
k_{\nu} k_{\mu} ) e_{\mu }  +
 i m ~\varepsilon_{\nu \mu\lambda\rho} k_{\mu} b_{\lambda\rho}= 0,~~~~~~~~~~\nn\\
(- k^{2} \eta_{\nu \mu } \eta_{\lambda\rho}
+k_{\nu} k_{\mu}\eta_{\lambda\rho}
-\eta_{\nu \mu } k_{\lambda} k_{\mu} ) b_{\mu\rho}  +
i {2 m \over 3}  ~\varepsilon_{\nu\lambda\mu \rho} k_{\mu} e_{\rho} = 0.
\eeqa
When $k^2 \neq M^2$ the system (\ref{systeminmomentum}) is off mass-shell and we have
four  pure gauge field solutions:
\beqa\label{puregauge}
e_{\mu}&=& k_{\mu},~~~~~~~b_{\nu\lambda} = 0;~~~~\nn\\
e_{\mu}&=&0,~~~~~~~~~ b_{\nu\lambda} =  k_{\nu}\xi_{\lambda}-k_{\lambda}\xi_{\nu}.
\eeqa
When $k^2 \neq M^2$ the system (\ref{systeminmomentum}) has seven solutions. These are four
pure gauge solutions (\ref{puregauge}) and additional three solutions representing
propagating modes:
\beqa\label{topphysicalmodes}
e^{(1)}_{\mu}= (0,1,0,0),~~
b_{\gamma\acute{\gamma}}^{(1)}={1\over i}{M \over \sqrt{\vec{k}^2 + M^2}}
\left(\begin{array}{cccc}
0&0&0&0 \\
0&0&0&0 \\
0&0&0&1 \\
0&0&-1&0 \\
\end{array} \right),
\nn\\
e^{(2)}_{\mu}= (0,0,1,0),~~
b_{\gamma\acute{\gamma}}^{(2)}=-{1\over i}{M \over \sqrt{\vec{k}^2 + M^2}}
\left(\begin{array}{cccc}
0&0&0&0 \\
0&0&0&1 \\
0&0&0&0\\
0&-1&0&0 \\
\end{array} \right),\nn\\
e^{(3)}_{\mu}= (0,0,0,{M \over \sqrt{\vec{k}^2 + M^2}}),~~
b_{\gamma\acute{\gamma}}^{(3)}={1\over i}
\left(\begin{array}{cccc}
0&0&0&0 \\
0&0&1&0\\
0&-1&0&0 \\
0&0&0&0 \\
\end{array} \right).
\eeqa
These propagating modes cannot be factorized into separately vector
or separately tensor solutions as it happens  for the pure
gauge solutions (\ref{puregauge}). It is a genuine superposition of
vector and tensor fields.
Let us consider the limit $M \rightarrow 0$. The above solutions will
factorize,  into two massless vector modes $ e^{(1)}_{\mu},~
e^{(2)}_{\mu},~$ of helicities $\lambda = \pm 1$
and helicity $\lambda = 0$ mode $b_{\gamma\acute{\gamma}}^{(3)}
$ of antisymmetric tensor - massless modes described in the previous section.
But when $M \neq 0$, in the rest
frame $\vec{k}^2 =0$, these solutions represent three spin-1 polarizations.

The above analysis suggests the following physical interpretation.
A massive spin-1  particle appears here as a vector field of helicities
$\lambda=\pm 1$ which acquires an extra polarization state absorbing
antisymmetric field of helicity $\lambda=0$, or as antisymmetric field of helicity  $\lambda=0$
which absorbs helicities $\lambda=\pm 1$ of the vector field.
It is sort of "dual" description of massive spin-1 particle.

In order to fully justify this phenomenon of superposition of polarizations
one should develop quantum-mechanical description of tensor fields \cite{Slavnov:1988sj,Allen:1990gb}.
It is a challenging problem
because in this field theoretical model we have infinite number of fields.
Not much is known about how to deal with such systems. One can develop "naive" Feynman
rules for transition amplitudes, but there is a need for deeper understanding
of the corresponding  path integral which is not only over  fields at
infinitely many space-time points, but also over infinitely many fields. It is impossible to make
truncation to finite number of fields without breaking their gauge symmetries.

\section{\it High-Rank Mass Terms}

Let us consider now the next invariant in five-dimensional space-time (4+1)
which can be constructed using totally antisymmetric Levi-Civita epsilon tensor
in combination with the generalized field strength tensors. It has the following form:
\beqa\label{ }
\Gamma_3~=~
\varepsilon_{\mu\nu\lambda\rho\sigma} Tr\{ G_{\mu\nu}G_{\lambda\rho, \sigma\alpha\alpha}
+ 2 G_{\mu\nu,\alpha} G_{\lambda\rho, \sigma\alpha} +
G_{\mu\nu,\sigma}G_{\lambda\rho, \alpha\alpha}\}.
\eeqa
As one can easily check this entity is also gauge invariant,
because under the gauge transformation
(\ref{fieldstrengthtensortransfor})  its variation vanishes:
$$
\delta_{\xi} \Gamma_3~=~0.
$$
It is not a metrically independent density, because not all Lorentz indices are contracted
by the epsilon tensor, part of them are contracted by the space-time metric. In this respect
it differs from the  density $\Gamma$, but it keeps other important properties of
density $\Gamma$ which we shall explore here. Indeed,
the density $\Gamma_3$ can be represented as a derivative of the  vector current:
\beqa
\Gamma_3~=~=
\partial_{\mu} \Xi_{\mu},~~~~~
\Xi_{\mu}
=  \varepsilon_{\mu\nu\lambda\rho\sigma} Tr \{G_{\nu\lambda }A_{\rho  \sigma\alpha\alpha}
+ 2 G_{\nu\lambda,\alpha }A_{\rho  \sigma\alpha}+
G_{\nu\lambda,  \alpha\alpha} A_{\rho\sigma }\}.
\eeqa
Considering the fifth component of the vector current $\Xi_{\mu}$
\beqa\label{topologicalcharge}
\Xi \equiv \Xi_{4}
&=&  \varepsilon_{4\nu\lambda\rho\sigma} Tr  \{G_{\nu\lambda }A_{\rho  \sigma\alpha\alpha}
+ 2 G_{\nu\lambda,\alpha }A_{\rho  \sigma\alpha}+G_{\nu\lambda,  \alpha\alpha} A_{\rho\sigma }\},
\eeqa
we shall reduce it to four dimensions\footnote{Here the index $\alpha$ can repeat the external index $\mu$.
Therefore we should separately consider the term
$ \varepsilon_{\nu\lambda\rho\sigma}
Tr  \{G_{\nu\lambda }A_{\rho  \sigma 4 4}
+ 2 G_{\nu\lambda,4 }A_{\rho  \sigma 4}+G_{\nu\lambda, 4 4} A_{\rho\sigma }\}$
as an additional expression in $\Xi_{4}$.
We have additional  tensor fields in four-dimensions:
$A_{\mu 4}=\tilde{A}_{\mu}, A_{\mu 44} = \tilde{\tilde{A}}_{\mu}, A_{\mu \nu 4} =\tilde{A}_{\mu \nu },
A_{\mu \nu 4 4}=\tilde{\tilde{A}}_{\mu \nu }$, and for them
the above invariant reduces to the form
$\varepsilon_{\nu\lambda\rho\sigma} Tr  \tilde{G}_{\nu\lambda }\tilde{A}_{\rho  \sigma  }$,
which we already studded  in the previous sections. In the following
consideration we shall take
all these additional fields equal to zero.}.
This is the case when the gauge fields are independent
on the fifth coordinate $x_4$. The density $\Xi $ is well
defined in four-dimensional space-time and it is gauge invariant
up to a total divergence term. Indeed,
its integral over the four-dimensional space-time\footnote{Below we are using the same
Greek letters to numerate now four-dimensional coordinates.}
\beqa\label{topologicalSigmaCS3}
\Xi
&=&  \varepsilon_{\nu\lambda\rho\sigma} \int  d^4 x ~Tr  \{G_{\nu\lambda }A_{\rho  \sigma\alpha\alpha}
+ 2 G_{\nu\lambda,\alpha }A_{\rho  \sigma\alpha}+G_{\nu\lambda,  \alpha\alpha} A_{\rho\sigma }\}
\eeqa
changes under the gauge variation  (\ref{polygauge}),
(\ref{fieldstrengthtensortransfor}) as follows
\beqa\label{variation}
\delta_{\xi} \Xi
&=& \varepsilon_{\nu\lambda\rho\sigma} \int Tr (
-i g [G_{\nu\lambda }~\xi] A_{\rho  \sigma\alpha\alpha}
+ G_{\nu\lambda } (\nabla_{\rho} \xi_{\sigma\alpha\alpha}
-i g  [A_{\rho  \sigma}~\xi_{\alpha\alpha}]  - 2i g [A_{\rho \alpha}~\xi_{ \sigma\alpha}] \nn\\
&-&2i g  [A_{\rho\sigma\alpha}~\xi_{\alpha}]- i g  [A_{\rho\alpha\alpha}~\xi_{\sigma}]
-i g  [A_{\rho  \sigma\alpha\alpha}~\xi])+
2(-i g [G_{\nu\lambda,\alpha }~\xi] -i g [G_{\nu\lambda }~\xi_{\alpha}])A_{\rho  \sigma\alpha}
\nn\\&+& 2G_{\nu\lambda,\alpha } (\nabla_{\rho} \xi_{\sigma\alpha}
-i g  [A_{\rho  \sigma}~\xi_{\alpha}]  - i g [A_{\rho \alpha}~\xi_{ \sigma}]
-i g  [A_{\rho  \sigma\alpha}~\xi])\nn\\
&+&
(-i g [G_{\nu\lambda,\alpha \alpha}~\xi] -2i g [G_{\nu\lambda \alpha}~\xi_{\alpha}]
-i g [G_{\nu\lambda }~\xi_{\alpha\alpha}])A_{\rho  \sigma}\nn\\
&+& G_{\nu\lambda,\alpha \alpha} (\nabla_{\rho} \xi_{\sigma}
-i g  [A_{\rho  \sigma}~\xi] )) d^4 x = \nn\\
&=&\varepsilon_{\nu\lambda\rho\sigma}
\int Tr  \partial_{\rho} (G_{\nu\lambda }   \xi_{\sigma\alpha\alpha}  +
2  G_{\nu\lambda, \alpha}   \xi_{\sigma\alpha}+
G_{\nu\lambda, \alpha\alpha}   \xi_{\sigma}) d^4 x  =0 ,\nn
\eeqa
and vanishes because terms in front of $\xi$, $\xi_{\alpha}$ and $\xi_{\alpha\alpha}$ cancel each other,
the others after integration by
part and recalling the Bianchi identities (\ref{newbianchi}), (\ref{newbianchi3})
reduce to the boundary terms which vanish  when
the gauge parameters $\xi_{\sigma\alpha\alpha}, \xi_{\sigma\alpha}$ and $ \xi_{\sigma}$
tend  to zero at infinity.

The dimension of this functional
is not difficult to calculate, in four dimensions the gauge fields have dimension of $[mass]^1$,
therefore if we intend to add this new term to the action we should introduce the next mass parameter $m_3$:
\beqa\label{topologicalmass3}
{m_3\over 2 }~\Xi
&=& {m_3\over 2 }~\varepsilon_{\nu\lambda\rho\sigma}  \int  Tr  \{ G_{\nu\lambda }A_{\rho  \sigma\alpha\alpha}
+ 2 G_{\nu\lambda,\alpha } A_{\rho  \sigma\alpha} +
 G_{\nu\lambda,\alpha\alpha} A_{\rho\sigma } \} d^4 x,
\eeqa
where $m_3$ has dimension of $[mass]^1$. To study the influence of this term on the
particle spectrum of the theory we have to consider quadratic on gauge
fields terms of the Lagrangian. The free equations of motion will take the following form\footnote{We
keep only the YM, rank-2 and rank-3 gauge fields in the free field equations.
The rank-4 and
high-rank gauge fields should be considered at subsequent levels.}:
\beqa
&&\partial_{\mu} F^{a}_{\mu\nu} +{1\over 2 }\partial_{\mu} (F^{a}_{\mu\nu,\lambda\lambda}
+ F^{a}_{\nu\lambda,\mu\lambda}
+ F^{a}_{\lambda\mu,\nu\lambda})+
{m \over 2} ~\varepsilon_{\nu \mu\lambda\rho} F^{a}_{\mu\lambda,\rho} = 0,\nn\\
&&\partial_{\mu} F^{a}_{\mu\nu,\lambda}
-{1\over 2} (\partial_{\mu} F^{a}_{\mu\lambda,\nu}
+\partial_{\mu} F^{a}_{\lambda\nu,\mu}
+\partial_{\lambda}F^{a}_{\mu\nu,\mu}
+\eta_{\nu\lambda} \partial_{\mu}F^{a}_{\mu\rho,\rho})\nn\\
&&+ {m \over 2} ~\varepsilon_{\nu\lambda\mu \rho} F^{a}_{\mu\rho} +
 {m_3 \over 2}  ~\varepsilon_{\nu\lambda \mu\rho} F^{a}_{\mu\rho,\sigma\sigma} +
 m_3  ~\varepsilon_{\nu \mu\rho\sigma} F^{a}_{\mu\rho,\sigma\lambda} = 0,\nn\\
&& \partial_{\mu}F^{a}_{\mu\nu,\lambda\rho}
-{1\over 3}\partial_{\mu}F^{a}_{\mu\lambda,\nu\rho}
- {1\over 3} \partial_{\mu}F^{a}_{\mu\rho,\nu\lambda}
+ {1\over 3}\partial_{\mu}F^{a}_{\nu\lambda,\mu\rho}
+{1\over 3} \partial_{\mu}F^{a}_{\nu\rho,\mu\lambda} +\\
&&+{1\over 3}\partial_{\lambda}F^{a}_{\nu\mu,\mu\rho}
+{1\over 3}\partial_{\rho}F^{a}_{\nu\mu,\mu\lambda}
+{1\over 6}\partial_{\lambda}F^{a}_{\nu\rho,\mu\mu}
+{1\over 6}\partial_{\rho}F^{a}_{\nu\lambda,\mu\mu}-\nn\\
&&-\eta_{\lambda\nu} ({1\over 3}\partial_{\mu}F^{a}_{\mu\sigma,\sigma\rho}
+{1\over 6}\partial_{\mu}F^{a}_{\mu\rho,\sigma\sigma} )
-\eta_{\nu\rho} ({1\over 3}\partial_{\mu}F^{a}_{\mu\sigma,\sigma\lambda}
+{1\over 6}\partial_{\mu}F^{a}_{\mu\lambda,\sigma\sigma} )+\nn\\
&&+ \eta_{\lambda\rho}({1\over 2} \partial_{\mu}F^{a}_{\mu\nu,\sigma\sigma}
- {1\over 3}\partial_{\mu}F^{a}_{\mu\sigma,\sigma\nu}
+{1\over 3}\partial_{\mu}F^{a}_{\nu\sigma,\sigma\mu})+\nn\\
&&+ \eta_{\lambda\rho}  \partial_{\mu}F^{a}_{\mu\nu }
- {1\over 2}( \eta_{\nu \rho } \partial_{\mu}F^{a}_{\mu\lambda}
+ \eta_{\nu\lambda }\partial_{\mu}F^{a}_{\mu\rho})+
{1\over 2}( \partial_{\rho}F^{a}_{\nu\lambda}
+ \partial_{\lambda}F^{a}_{\nu\rho})+\nn\\
&&+ {m_3 \over 2} (~\varepsilon_{\nu \lambda\mu\sigma} F^{a}_{\mu\sigma,\rho}
+  \varepsilon_{\nu \rho\mu\sigma} F^{a}_{\mu\sigma,\lambda}  +
 \varepsilon_{\nu \mu \gamma \sigma } \eta_{\lambda\rho} F^{a}_{\mu\gamma,\sigma}  )=
0.\nn
\eeqa
From the second equation for the rank-2 gauge field it follows that
now its symmetric part $A^{S}_{\nu\lambda}$  interacts through the
second mass term with the antisymmetric part of the rank-3 gauge field
$$
{m_3 \over 2} ~(\varepsilon_{\nu \mu\rho\sigma} F^{a}_{\mu\rho,\sigma\lambda}+
\varepsilon_{\lambda\mu\rho\sigma} F^{a}_{\mu\rho,\sigma\nu })=
 m_3  ~(\varepsilon_{\nu \mu\rho\sigma} \partial_{\mu} A^{a}_{ \rho \sigma\lambda}+
\varepsilon_{\lambda\mu\rho\sigma} \partial_{\mu} A^{a}_{ \rho \sigma\nu })
$$
and from the third equation - that the rank-3 gauge field interacts with the rank-2 gauge field,
so that together they form a coupled system of equations similar to the one considered in the previous section
and can produce massive particle of spin-2.
In general it is a complicated system of coupled linear equations and full understanding
of its solutions requires detailed analysis which we shall provide elsewhere.

At the end of this section we shall present the general form of the invariant which can be
constructed in terms of higher-rank field strength tensors and epsilon tensor in four
dimensions
\beqa\label{topologicalSigmaCSGeneral}
\Sigma_{2s+1}
&=&  \varepsilon_{\mu\nu\rho\sigma} \int  d^4 x ~Tr  \{
G_{\mu\nu}A_{\rho  \sigma \lambda_{i_1}\lambda_{i_1} ...\lambda_{i_s}\lambda_{i_s}}
+ ... +G_{\mu\nu,\lambda_{i_1}\lambda_{i_1} ...\lambda_{i_s}\lambda_{i_s}} A_{\rho\sigma }\}.
\eeqa
As we already suggested, it can be added to the massless Lagrangian (\ref{fulllagrangian3}) with
different mass parameters $m_{2s+1}$ as in (\ref{fulllagrangiantop}). The consequences of this extension on the
particle spectrum is not so easy to analyze and some general
method should be developed to resolve the particle spectrum at subsequent levels.

\section{\it Topological Density in Six Dimensions}

In the previous sections we considered the densities in five and four dimensions.
It is also possible to construct invariants in higher dimensions. First let us
consider metric-independent  density in six dimensions:
\beqa\label{invarinatinsixdimensions}
\Delta
= \varepsilon_{\mu\nu\lambda\rho\sigma\kappa} ~Tr ~G_{\mu\nu,\lambda}G_{\rho\sigma,\kappa },
\eeqa
which is  a gauge invariant entity, because under  transformation $\delta_{\xi} $
(\ref{fieldstrengthtensortransfor}) it vanishes:
\beqa
\delta_{\xi} \Delta  &=& \varepsilon_{\mu\nu\lambda\rho\sigma\kappa}
Tr (\delta G_{\mu\nu,\lambda} G_{\rho\sigma, \kappa}+
G_{\mu\nu,\lambda} \delta G_{\rho\sigma, \kappa}) \nn\\
&=&-i g  \varepsilon_{\mu\nu\lambda\rho\sigma\kappa}
Tr ( [G_{\mu\nu,\lambda} ~\xi] +[G_{\mu\nu} ~\xi_{\lambda}]) G_{\rho\sigma,\kappa}
 +G_{\mu\nu,\lambda} (~[~G_{\rho\sigma,\kappa}~ \xi ]
+  [G_{\varrho\sigma} ~\xi_{\kappa}]~))=0.\nn
\eeqa
The $\Delta $ is obviously diffeomorphism-invariant and does not involve a space-time metric.
It is also true that $\Delta $ is a total derivative of a
vector current $\Pi_{\mu}$. Indeed, simple algebraic computation gives
\beqa\label{exactformhigh}
\Delta
=\varepsilon_{\mu\nu\lambda\rho\sigma\kappa} ~
Tr ~G_{\mu\nu,\lambda}G_{\rho\sigma, \kappa}=
2 ~\partial_{\mu} \Pi_{\mu},
\eeqa
where
\beqa\label{topologicalcurrenthigh}
\Pi_{\mu}
&=&   \varepsilon_{\mu\nu\lambda\rho\sigma\kappa} ~Tr ~ G_{\nu\lambda,\rho }  A_{\sigma \kappa }.
\eeqa
While the invariant $\Delta $ and the vector current $\Pi_{\mu}$ are defined on a six-dimensional
manifold, we may restrict the latter to one lower, five-dimensional manifold.
Considering the sixth component of the vector current $\Pi_{\mu}$
\beqa\label{topologicalchargehigh}
\Pi \equiv \Pi_{5}
&=&  \varepsilon_{5\nu\lambda\rho\sigma\kappa} Tr G_{\nu\lambda,\rho   }A_{\sigma \kappa}.
\eeqa
one can see that the remaining indices will not repeat
the external index and the sum is restricted to the sum over indices of five-dimensional
space-time.
Therefore we can reduce this functional to five dimensions.
This is the case when the gauge fields are independent
on the sixth coordinate $x_5$. Thus the density $\Pi$ is well
defined in five-dimensional space-time and, as we shall see,
it is also gauge invariant up to the total divergence term.  Therefore we shall
consider its integral over five-dimensional space-time:
\beqa\label{topologicalSigmaCShigh}
\int_{M_5} d^5 x~ \Pi
&=&  \varepsilon_{\nu\lambda\rho\sigma\kappa} \int_{M_5} d^5 x~ Tr~ G_{\nu\lambda, \rho  } A_{\sigma\kappa} .
\eeqa
This functional is gauge invariant up to the total divergence term.
Its gauge variation under $\delta_\xi$ (\ref{polygauge}),
(\ref{fieldstrengthtensortransfor}) is
\beqa\label{variationhigh}
\delta_{\xi} \int_{M_5} d^5 x~ \Pi
=\varepsilon_{\nu\lambda\rho\sigma\kappa} \int_{M_5} \partial_{\sigma} ~Tr (
G_{\nu\lambda,\rho }  ~\xi_{\kappa} ) d^5 x=\varepsilon_{\nu\lambda\rho\sigma\kappa}
\int_{\partial M_5} Tr (
G_{\nu\lambda, \rho}  \xi_{\kappa} ) d\sigma_{\sigma} =0 ,\nn
\eeqa
where the boundary term vanishes  when
the gauge parameter $\xi_{\kappa}$ tends to zero at infinity.

One can  construct higher-rank extension of the above densities.
The first in this infinite series is
\beqa\label{topologicalsixdim4}
\Delta_{3}
&=&  \varepsilon_{\mu\nu\lambda\rho\sigma\kappa}
G_{\mu\nu,\lambda\alpha} G_{\rho\sigma,\kappa\alpha } ~~,
\eeqa
the second has two varieties
\beqa\label{topologicalsixdim5}
\Delta_{5}
&=&  \varepsilon_{\mu\nu\lambda\rho\sigma\kappa}
G_{\mu\nu,\lambda\alpha\beta} G_{\rho\sigma,\kappa\alpha\beta }~, \nn\\
\Delta^{'}_{5}
&=&  \varepsilon_{\mu\nu\lambda\rho\sigma\kappa}
G_{\mu\nu,\lambda\alpha\alpha} G_{\rho\sigma,\kappa\beta\beta } \nn\\
\eeqa
and so on. These invariants may be important for the physics of
D-branes in corresponding dimensions.

\section{\it Note Added}

In the higher-spin literature
and, in particular, in the work of Metsaev and others,
it was shown that the consistency of the interaction vertices with the Poincar\'e symmetry
requires that the cubic interaction vertices should contain a number of derivatives greater or
equal to $s_1 + s_2 + s_3 - 2 s_{min}$, where $s_a, a=1,2,3$ are the spins of the interacting particles.
This result seems to be in a contrast with the form of the interaction
vertices in the generalized Yang-Mills theory \cite{Savvidy:2005fi,Savvidy:2005zm,Savvidy:2005ki},
 in which all interaction vertices
between high-spin fields have dimensionless coupling constants in four-dimensional space-time.
That is, {\it the cubic interaction vertices have only first order derivatives, there is no self-interaction
cubic vertices and that the quartic vertices have no derivatives}.

Let us see first why the generalized Yang-Mills theory predicts that the cubic interaction vertices have
only first order derivatives and that it avoids self-interaction cubic vertices.
The general structure of the vertices is defined by the {\it gauge and Lorentz invariant Lagrangian}
(\ref{fulllagrangiantop}), which is quadratic in the field strength tensors (\ref{fulllagrangian1}) and (\ref{secondfulllagrangian}).
The field strength tensors themselves are quadratic in high-spin fields (\ref{fieldstrengthparticular}),
therefore the Lagrangian (\ref{fulllagrangiantop}) contains only quadratic, cubic and quartic vertices.
The cubic vertices appear in the product of the derivative terms and quadratic terms of the field
strength tensors (\ref{fieldstrengthparticular}) and have the following general structure:
$g \partial A_{s_1} A_{s_2} A_{s_3} $, while there is no self-interaction vertices with $s_1=s_2=s_3$.
The quartic vertices appear in the product of quadratic terms
$g^2 A_{s_1} A_{s_2}A_{s_3} A_{s_4}$. These structure of the vertices is a consequence of the gauge
and Lorentz invariance of the Lagrangian (\ref{fulllagrangiantop}), which has been   proven explicitly
by using formulas (\ref{polygauge}) and (\ref{fieldstrengthtensortransfor}) \cite{Savvidy:2005fi,Savvidy:2005zm,Savvidy:2005ki}.

Next, let us see that: \\
A) from the work of the G\"oteborg group it also
follows that there exists a large class of Poincar\'e invariant cubic vertices for high spin fields
which have only first order derivatives, in agreement with the cubic vertices of the generalized
Yang-Mills theory,\\
B) in the work of Metsaev there is a place where the author has made
assumption  which leads him to the conclusion that the cubic
vertices should have higher derivatives, in contrast with the results of G\"oteborg group,\\
C) from spinor representation of the cubic vertices for high spin fields it also
follows that there are dimensionless Poincar\'e invariant cubic vertices
\cite{Benincasa:2007xk,Georgiou:2010mf}.\\
D) dimensionless cubic vertices for high spin fields appear in open string theory
with Chan-Paton charges when one compute tree-level scattering amplitudes
\cite{Savvidy:2008ks,Taronna:2010qq,Antoniadis:2009rd,Schlotterer:2010kk}.
\\

A) Let us review the results of the G\"oteborg group which are published in
\cite{Bengtsson:1983pd,Bengtsson:1983pg,Bengtsson:1986kh}.
In the light-front formulation of relativistic dynamics used
in \cite{Bengtsson:1983pd,Bengtsson:1983pg,Bengtsson:1986kh} the massless
particles of spin-s are described by a
complex function $\phi_s$ which encodes two physical helicities $h=\pm s$ of the massless particles.
In this approach there are no auxiliary fields and
questions associated with the gauge invariance, because it permits to work with the physical fields
$\phi_s$ exclusively. What one should
be concerned of is the relativistic invariance of the scattering amplitudes. The Poincar\'e group
is realized here non-linearly and one should derive the
self-interaction cubic vertices  as a non-linear realization of the Poincar\'e group
\cite{Bengtsson:1983pd,Bengtsson:1983pg}. The authors came to the
conclusion that there are $s$ derivatives in the cubic self-interaction vertices and
the coupling constant has dimension of $[mass]^{1 -s}$.
This result of Lars Brink and his collaborators
raised expectations that a consistent interacting theory might exist in flat space-time, because
this approach demonstrated the existence of physically non-trivial interaction of high spin particles.

The next important step has been made in \cite{Bengtsson:1986kh} where the authors derived
the  cubic vertices for all massless bosonic representations of the Poincar\'e group
which includes {\it interactions between different spins} $s_1,s_2,s_3$.
For the cubic vertices  in four dimensions they found the following
expressions (see formulas (A1.4-6) in \cite{Bengtsson:1986kh}):
$$
M_3 = \int D \beta^{s_1}_{1} \beta^{-s_2}_{2} \beta^{-s_3}_{3}
\bar{P}^{(s_2 + s_3 -s_1)} \bar{\phi}_{s_1}(1)
\phi_{s_2}(2) \phi_{s_3}(3) + CC,~~~~if~~~~~~s_2 + s_3 > s_1,  ~~~~~ (A1.4)
$$
$$
M_3 = \int D \beta^{-s_1}_{1} \beta^{s_2}_{2} \beta^{s_3}_{3}
\bar{P}^{(s_1 -s_2 -s_3)}  \phi_{s_1}(1)
\bar{\phi}_{s_2}(2) \bar{\phi}_{s_3}(3) +CC,~~~~if~~~~~~s_1 > s_2 + s_3,  ~~~~~ (A1.5)
$$
$$
M_3 = \int D \beta^{-s_1}_{1} \beta^{-s_2}_{2} \beta^{-s_3}_{3}
\bar{P}^{(s_1 +s_2 +s_3)}  \phi_{s_1}(1)
 \phi_{s_2}(2)  \phi_{s_3}(3) +CC,   ~~~~~~~~~~~~~~~~~~~~~~~~~~~~~~~~~~~ (A1.6)
$$
where $\beta_a =  2 p^+_{a}$ and $a=1,2,3$ numerates the interacting particles, D denotes the momentum
integration and momentum delta functions.  The transverse momentum is
$P =  {1\over 3}\sum_a  \hat{\beta}_a p_a, ~~
\hat{\beta}_a = \beta_{a+1} -\beta_{a+2}$. Here transverse momenta $p_a$, fields $\phi_s$
and all vectors  $A$ are defined as complex variables $A=A_1 +iA_2,~ \bar{A}=A_1 -i A_2$.
If one takes the spins of the scattered particles such that
\be\label{restrictions}
s_2 + s_3 - s_1 =1,~~~~or~~~~~~~s_1 - s_2 - s_3 =1,,~~~~or~~~~~~~s_1 + s_2 + s_3 =1,
\ee
then one can get the {\it cubic vertices  (A1.4-6) which are linear in momentum $P$},
in agreement with the cubic vertices of the generalized
Yang-Mills theory \cite{Savvidy:2005fi,Savvidy:2005zm,Savvidy:2005ki}.

B) It seems to me that the following assumption in the work of Metsaev \cite{Metsaev:2005ar,Metsaev:2007rn}, who follows
the light-front formulation of the G\"oteborg group,  is not necessary and
therefore leads him to a different conclusion. His general formula for the cubic vertex (5.9)
in \cite{Metsaev:2005ar} is
$$
M_3 \sim Z^{(s_1+s_2+s_3 -k)/2} \prod B^{s_a + (k-s_1-s_2-s_3)/2}_a,~~~~~~~~~~~~~~~(5.9)
$$
where $Z$ and $B_a, a=1,2,4$ are linear in momenta functions and the vertex has therefore k powers
of the transverse momenta.
It is assumed that:  "The powers of the forms $B_a$ and $Z$ in (5.9) must be non–negative integers." This
leads the author to the
conclusion that the number of derivatives in the vertex should be greater or equal to $s_1 + s_2 + s_3 - 2 s_{min}$.
But as it was demonstrated in \cite{Bengtsson:1986kh}  the ratio of two polynomials of momenta can be
reducible in four dimensions, therefore  one should allow negative integer powers as well.
This leads the authors of \cite{Bengtsson:1986kh} to the vertices (A1.4-6).

C) As it was demonstrated in
\cite{Bengtsson:1986kh,Berends:1984rq,Berends:1985xx,Berends:1984wp,Manvelyan:2010je},
the "morphology" of the available
invariant vertices is much richer when the interaction between different spins
is allowed (\ref{restrictions}). The main difficulty here is to
derive or to guess the genuine form of the full Lagrangian which is behind
the perturbative constructions, that is, to extend the results to the second and higher orders
in the deformation parameter. There is a need here to understand  the structure of high spin interactions
beyond the perturbation theory. The spinor representation of the scattering
amplitudes may offer such a solution. One can get the dimensionless cubic vertices using
the results of the Benincasa and Cachazo \cite{Benincasa:2007xk}, they  are
\cite{Georgiou:2010mf}:
\beqa\label{dimensionone1}
&M_3 = f <1,2>^{-2h_1 -2h_2 -1} <2,3>^{2h_1 +1} <3,1>^{2h_2 +1},~~~~&h_3= -1 - h_1 -h_2, \nn\\
&M_3 = k ~[1,2]^{2h_1 +2h_2 -1} ~[2,3]^{-2h_1 +1} ~[3,1]^{-2h_2 +1},~~~~~~~~~~~~~~~&h_3= 1 - h_1 -h_2.
\eeqa
and are identical to the conditions (\ref{restrictions}).
The formulas (\ref{dimensionone1}) give a general expression for the cubic vertices  in terms
of two independent helicities $h_1$ and $h_2$. It allows to choose any
$h_1$ and $h_2$ and then to find out $h_3$ for which the three-particle interaction
vertex in four-dimensional space-time will have dimensionless coupling constants f and h.
The details can be found in \cite{Benincasa:2007xk,Georgiou:2010mf}.

\section{\it Appendix}

The extended non-Abelian gauge transformation $
\delta_{\xi} $ of the tensor gauge fields
is defined
by the equations \cite{Savvidy:2005fi,Savvidy:2005zm,Savvidy:2005ki}:
\beqa\label{polygauge}
\delta_{\xi}  A_{\mu} &=& \partial_{\mu}\xi -i g[A_{\mu},\xi]\nonumber\\
\delta_{\xi}  A_{\mu\nu} &=& \partial_{\mu}\xi_{\nu} -i g[A_{\mu},\xi_{\nu}]
-i g [A_{\mu\nu},\xi]\nonumber\\
\delta_{\xi}  A_{\mu\nu\lambda} &=& \partial_{\mu}\xi_{\nu\lambda}
-i g[A_{\mu},\xi_{\nu\lambda}]-
i g[A_{\mu\nu},\xi_{\lambda}]-i g [A_{\mu\lambda},\xi_{\nu}]
-i g [A_{\mu\nu\lambda},\xi],\\
&~&..............................,\nn
\eeqa
where $\xi^{a}_{\lambda_1 ... \lambda_{s}}(x)$ are totally symmetric gauge parameters,
and comprises a closed algebraic structure.
The tensor gauge fields are in the matrix representation
$A^{ab}_{\mu\lambda_1 ... \lambda_{s}} =
(L_c)^{ab}  A^{c}_{\mu\lambda_1 ... \lambda_{s}} = i f^{acb}A^{c}_{\mu
\lambda_1 ... \lambda_{s}}$  and
$f^{abc}$ are the structure constants. The generalized field strength tensors
are defined as follows
\cite{Savvidy:2005fi,Savvidy:2005zm,Savvidy:2005ki}:
\beqa\label{fieldstrengthparticular3}
G_{\mu\nu} &=&
\partial_{\mu} A_{\nu} - \partial_{\nu} A_{\mu} -
i g [A_{\mu}~A_{\nu}],\\
G_{\mu\nu,\lambda} &=&
\partial_{\mu} A_{\nu\lambda} - \partial_{\nu} A_{\mu\lambda} -
i g  (~[A_{\mu}~A_{\nu\lambda}] + [A_{\mu\lambda}~A_{\nu}] ~),\nn\\
G_{\mu\nu,\lambda\rho} &=&
\partial_{\mu} A_{\nu\lambda\rho} - \partial_{\nu} A_{\mu\lambda\rho} -
i g (~[A_{\mu}~A_{\nu\lambda\rho}] +
 [A_{\mu\lambda}~A_{\nu\rho}]+[A_{\mu\rho}~A_{\nu\lambda}]
 + [A_{\mu\lambda\rho}~A_{\nu}] ~),\nn\\
 ......&.&............................................\nn
\eeqa
and  transform homogeneously
with respect to the extended gauge transformations $\delta_{\xi} $:
\beqa\label{fieldstrengthtensortransfor}
\delta G^{a}_{\mu\nu}&=&  -i g  [G_{\mu\nu} ~\xi] , \\
\delta G^{a}_{\mu\nu,\lambda} &=& -i g  (~[~G_{\mu\nu,\lambda}~ \xi ]
+  [G_{\mu\nu} ~\xi_{\lambda}]~),\nonumber\\
\delta G^{a}_{\mu\nu,\lambda\rho} &=& - i g
(~[G^{b}_{\mu\nu,\lambda\rho} ~\xi]
+ [ G_{\mu\nu,\lambda} ~\xi_{\rho}] +
[G_{\mu\nu,\rho} ~\xi_{\lambda}] +
[G_{\mu\nu} ~\xi_{\lambda\rho}]~),\nn\\
 ......&.&............................................\nn
\eeqa
The field strength tensors fulfil the Bianchi
identities \cite{Konitopoulos:2007hw}.
In the YM theory the Bianchi identity is
\be
[\nabla_{\mu},G_{\nu\lambda}]+[\nabla_{\nu},G_{\lambda\mu}]+
[\nabla_{\lambda},G_{\mu\nu}]=0,
\ee
and for the higher-rank field strength tensors $G_{\nu\lambda,\rho}$ and
$G_{\nu\lambda,\rho\sigma}$ they are:
\be\label{newbianchi}
[\nabla_{\mu},G_{\nu\lambda,\rho}]-ig[A_{\mu\rho},G_{\nu\lambda}]+
[\nabla_{\nu},G_{\lambda\mu,\rho}]-ig[A_{\nu\rho},G_{\lambda\mu}]+
[\nabla_{\lambda},G_{\mu\nu,\rho}]-ig[A_{\lambda\rho},G_{\mu\nu}]=0,
\ee
\be\label{newbianchi3}
[\nabla_{\mu},G_{\nu\lambda,\rho\sigma}]-ig[A_{\mu\rho},G_{\nu\lambda,\sigma}]
-ig[A_{\mu\sigma},G_{\nu\lambda,\rho}]-ig[A_{\mu\rho\sigma},G_{\nu\lambda}]
+ cyc.perm. (\mu\nu\lambda)=0
\ee
and so on.

\vfill
\end{document}